%% file: paper.tex
\documentclass[preprint]{sigplanconf}
\pdfoutput=1

\usepackage{natbib}
\usepackage[disable]{todonotes}
\usepackage{xcolor}
\usepackage{paralist}
\usepackage{verbatim}
\usepackage{multirow}
\usepackage{graphicx}
\usepackage{listings, url}
\usepackage{amssymb}
\usepackage{color}
\usepackage{subcaption}
\balancecolumns

\definecolor{mygreen}{rgb}{0,0.6,0}
\definecolor{mygray}{rgb}{0.5,0.5,0.5}
\definecolor{mymauve}{rgb}{0.58,0,0.82}

\begin{document}

\title{A Programming Model and Runtime System for Significance-Aware Energy-Efficient Computing}
\authorinfo{Vassilis Vassiliadis$^1$ \and Konstantinos Parasyris$^2$ \and Charalambos Chalios$^3$ \and Christos D. Antonopoulos$^4$ \and Spyros Lalis$^5$ \and Nikolaos Bellas$^6$ \and Hans Vandierendonck$^7$ \and Dimitrios S. Nikolopoulos$^8$}
{\begin{tabular}{ccc}
$^{1,2,4,5,6}$Electrical and Computer Eng. Dept. & $^{1,2,4,5,6}$Centre for Research and Technology Hellas & $^{3,7,8}$Queen's University Belfast\\
University of Thessaly, Greece & (CE.R.T.H.), Greece & United Kingdom
\end{tabular}
}
{\{vasiliad$^1$,koparasy$^2$,cda$^4$,lalis$^5$,nbellas$^6$\}@uth.gr \and \{cchalios01$^3$,h.vandierendonck$^7$,d.nikolopoulos$^8$\}@qub.ac.uk}

\maketitle

\begin{abstract}
Reducing energy consumption is one of the key challenges in computing technology. One factor that contributes to high energy consumption is that all parts of the program are considered equally significant for the accuracy of the end-result. However, in many cases, parts of computations can be performed in an approximate way, or even dropped, without affecting the quality of the final output to a significant degree. 

In this paper, we introduce a task-based programming model and runtime system that exploit this observation to trade off the quality of program outputs for increased energy-efficiency. This is done in a structured and flexible way, allowing for easy exploitation of different execution points in the quality/energy space, without code modifications and without adversely affecting application performance. The programmer specifies the significance of tasks, and optionally provides approximations for them. Moreover, she provides hints to the runtime on the percentage of tasks that should be executed accurately in order to reach the target quality of results. The runtime system can apply a number of different policies to decide whether it will execute each individual less-significant task in its accurate form, or in its approximate version. Policies differ in terms of their runtime overhead but also the degree to which they manage to execute tasks according to the programmer's specification. 

The results from experiments performed on top of an Intel-based multicore/multiprocessor platform show that, depending on the runtime policy used, our system can achieve an energy reduction of up to 83\% compared with a fully accurate execution and up to 35\% compared with an approximate version employing loop perforation. At the same time, our approach always results in graceful quality degradation.

\end{abstract}


\input{introduction}
\input{progr_model}
\input{runtime}
\input{experiments}
\input{related}

\input{conclusions}

\bibliographystyle{abbrvnat}
\bibliography{ppopp15}

\end{document}

%% file: introduction.tex
\section{Introduction}

Energy consumption has become a major barrier, not only for tetherless computing -- the traditional energy-constrained environment -- but also for other computing domains, including big science. Building an exascale machine with today's technology is impractical due to the inordinate power draw it would require, hampering large-scale scientific efforts. Likewise, current technologies are too energy-inefficient to realize smaller and smarter embedded/wearable devices for a wide range of ubiquitous computing applications that can greatly benefit society, such as personalized health systems.

One factor that contributes to the energy footprint of current computer technology is that all parts of the program are considered to be equally important, and thus are all executed with full accuracy. However, as shown by previous work on approximate computing, in several classes of computations, not all parts or execution phases of a program affect the quality of its output equivalently. In fact, the output may remain virtually unaffected even if some computations produce incorrect results or fail completely. Data intensive applications and kernels from multimedia, data mining, and visualization algorithms, can all tolerate a certain degree of imprecision in their computations. For example, Discrete Cosine Transform (DCT), a module of popular video compression kernels, which transforms a block of image pixels to a block of frequency coefficients, can be partitioned into layers of significance, owing to the fact that human eye is more sensitive to lower spatial frequencies, rather than higher ones. By explicitly tagging operations that contribute to the computation of higher frequencies as less-significant, one can leverage smart underlying system software to trade-off video quality with energy and performance improvements.

In this paper, we introduce a novel, significance-driven programming environment for approximate computing, comprising a programming model, compilation toolchain and runtime system. The environment allows programmers to trade-off the quality of program outputs for increased energy-efficiency, in a structured and flexible way. The programming model follows a task-based approach. For each task, the developer declares its significance depending on how strongly the task contributes to the quality of the final program output, and provides an approximate version of lower complexity that returns a less accurate result or just a meaningful default value. Also, the developer controls the degradation of output quality, by specifying the percentage of tasks to be executed accurately. In turn, the runtime system executes tasks on available cores in a significance-aware fashion, by employing the approximate versions of less-significant tasks, or dropping such tasks altogether. This can lead to shorter makespans and thus to more energy-efficient executions, without having a significant impact on the results of the computation. 

The main contributions of this paper are the following: 
\begin{inparaenum}[(i)]
\item We propose a new programming model that allows the developer to structure the computation in terms of distinct tasks with different levels of significance, to supply approximate task versions, and to control the degradation of program outputs; 
\item We introduce different runtime policies for deciding which tasks to execute accurately to meet the programmer's specification; 
\item We implement compiler and runtime support for the programming model and the runtime policies. 
\item We experimentally evaluate the potential of our approach, as well as the performance of the runtime policies.
\end{inparaenum} 

Previous work has already explored the potential of approximate computing for specific algorithms and software blocks. Our work is largely complementary to these efforts, as we introduce a programming model that makes it possible to apply such techniques in task-based programs that can exploit the parallelism of modern many-core platforms. There are also major differences with other approximate computing frameworks. For instance, the granularity of approximation is at the level of tasks, rather than individual data types, variables or arithmetic operations.  Our programming model operates not only at a different granularity but also at a different level of abstraction for approximate computing --relative significance of code blocks--, which enables the compiler and runtime system to implement different policies that trade energy savings with quality. Also, one can explore different points in the quality/energy space in an easy and direct way, without code modifications, simply by specifying the percentage of tasks that should be executed accurately - this can be an open parameter of a kernel or an entire application, which can take different values in each invocation, or be changed interactively by the user. 

The rest of the paper is structured as follows. Section~\ref{sec:Progr_Model} introduces the programming model. Section~\ref{sec:runtime} discusses the runtime system, and the different policies used to drive task execution. Section~\ref{sec:eval} presents the experimental evaluation on top of an Intel-based 16-way multiprocessor/multicore platform, using a set of benchmark kernels that were ported to our programming model. Section~\ref{sec:related_work} gives an overview of related work. Finally, Section~\ref{sec:conclusions} concludes the paper and identifies directions for future work.

%% file: progr_model.tex
\section{Programming Model\label{sec:Progr_Model}}

Improving energy consumption by controllably reducing the quality of application output has been already identified as an attractive option in the domain of power-sensitive HPC programming. Part of the problem of energy inefficiency is that all computations are treated as equally important, despite the fact that only a subset of these computations may be critical in order to achieve an acceptable quality of service (QoS). A key challenge though is how to identify and tag computations of the program which must be executed accurately from those that are of less importance and thus can be executed approximately.

In this section we introduce a programming model that allows the programmer to express her perspective on the significance of the contribution of each computation to the quality of the final output.Highly significant computations are executed accurately, whereas non-significant computations can be executed approximately, at the expense of errors, or can be totally dropped.

Our vision is to elevate significance characterization as a first class concern in software development, similar to parallelism and other algorithmic properties traditionally being in the focus of programmers. To this end, the main objectives of the proposed programming model are the following:
\begin{itemize}
    \item to allow programmers to express the significance of computations in terms of their contribution to the quality of the end-result; 
    \item to allow programmers to specify approximate alternatives for selected computations;
    \item to allow programmers to express parallelism, beyond significance;
    \item to allow programmers to control the balance between energy consumption and the quality of the end-result, without sacrificing performance;
    \item to enable optimization and easy exploration of trade-offs at execution time;
    \item to be user friendly and architecture agnostic.
\end{itemize}

Programmers express significance semantics using {\it \#pragma} compiler directives. Pragmas-based programming models facilitate non-invasive and progressive code transformations, without requiring a complete code rewrite. We adopt a task-based paradigm, similarly to OmpSS~\cite{duran2011ompss} and the latest version of OpenMP~\cite{openmp4_2013}. Task-based models offer a straightforward way to express communication across tasks, by explicitly defining inter-task data dependencies. Parallelism is expressed by the programmer in the form of independent tasks, however the scheduling of the tasks is not explicitly controlled by the programmer, but is performed at runtime, also taking into account the data dependencies among tasks. 


\begin{figure}[htb]
\begin{lstlisting}[caption={Programming model use case: Sobel filter}, label={src:sobel}, language=C, basicstyle=\ttfamily\scriptsize]
int sblX(const unsigned char img[], int y, int x) {
  return img[(y-1)*WIDTH+x-1]
    + 2*img[y*WIDTH+x-1] + img[(y+1)*WIDTH+x-1]
    - img[(y-1)*WIDTH+x+1]
    - 2*img[y*WIDTH+x+1] - img[(y+1)*WIDTH+x+1];
}

int sblX_appr(const unsigned char img[], 
              int y, int x) {
  return /* img[(y-1)*WIDTH+x-1]   Ommited taps */<@{\label{src:sbltaskapprskip1}}@>
    + 2*img[y*WIDTH+x-1] + img[(y+1)*WIDTH+x-1]
    /* - img[(y-1)*WIDTH+x+1]   Ommited taps *//<@{\label{src:sbltaskapprskip2}}@>
    - 2*img[y*WIDTH+x+1] - img[(y+1)*WIDTH+x+1];
}

/* sblY and sblY_appr are similar */

void sbl_task(unsigned char res[], 
              const unsigned char img[], int i) {
  unsigned int p, j;

  for (j=1; j<WIDTH-1; j++) {
    p = sqrt(pow(sblX(img, i, j),2) + 
             pow(sblY(img, i, j),2));
    res[i*WIDTH + j] = (p > 255) ? 255 : p;
  }
}

void sbl_task_appr(unsigned char res[], <@{\label{src:sbltaskapprbegin}}@>
                  const unsigned char img[], int i) {
  unsigned int p, j;

  for (j=1; j<WIDTH-1; j++) {
    /* abs instead of pow/sqrt, 
       approximate versions of sblX, sblY */
    p = abs(sblX_appr(img, i, j) + <@{\label{src:sbltaskapprabs}}@>
            sblY_appr(img, i, j));
    res[i*WIDTH + j] = (p > 255) ? 255 : p;  
  }
}<@{\label{src:sbltaskapprend}}@>

double sobel(void) {
  int i;
  unsigned char img[WIDTH*HEIGHT], res[WIDTH*HEIGHT];

  /* Initialize img array and reset res array */
  ...
  for (i=1; i<HEIGHT-1; i++) <@{\label{src:mainforbegin}}@>
    <@{\bfseries\color{blue}\#pragma omp task label}@>(sobel) <@{\bfseries\color{blue}in}@>(img) <@{\bfseries\color{blue}out}@>(res) \<@{\label{src:taskinoutlabel}}@>
    <@{\bfseries\color{blue}significant}@>((i%9 + 1)/10.0) <@{\bfseries\color{blue}approxfun}@>(sbl_task_appr)<@{\label{src:tasksig}}@>
      sbl_task(res, img, i); /* Compute a single 
                                output image row */
  }<@{\label{src:mainforend}}@>
  <@{\bfseries\color{blue}\#pragma omp taskwait label}@>(sobel) <@{\bfseries\color{blue}ratio}@>(0.35)<@{\label{src:sobeltaskwait}}@>
}

\end{lstlisting}
\end{figure}

Listing~\ref{src:sobel} illustrates the use of our programming model, using the Sobel filter as a running example.

\begin{center}
\begin{minipage}{\columnwidth}
\begin{lstlisting}[caption={\#pragma omp task}, label={src:omptask}]
#pragma omp task [significant(expr(...))]
  [approxfun(function())]
  [label(...)] [in(...)] [out(...)] 
\end{lstlisting}
\end{minipage}
\end{center}

Tasks are specified using the {\it \#pragma omp task} directive (Listing~\ref{src:omptask}), followed by a function which is equivalent to the task body. 

The significance of the task is specified through the {\it significant()} clause.  
Significance takes values in the range [0.0, 1.0] and characterizes the relative importance of tasks for the quality of the end-result of the application. Depending on their (relative) significance, tasks may be approximated or dropped at runtime. The special values 1.0 and 0.0 are used for tasks that must {\it unconditionally} be executed accurately and approximately, respectively.

For tasks with significance less than 1.0, the programmer may provide an alternative, approximate task body, through the {\it approxfun()} clause. This function is executed whenever the runtime opts for a non-accurate computation of the task. It typically implements a simpler, approximate version of the computation, which may even degenerate to just setting default values to the output. If a task is selected by the runtime system to be executed approximately, and the programmer has not supplied an {\it approxfun} version, it is simply dropped by the runtime. It should be noted that the {\it approxfun} function implicitly takes the same arguments as the function implementing the accurate version of the task body.

Programmers explicitly specify data flow to the task through the {\it in()} and {\it out()} clauses. This information is exploited by the runtime to automatically determine the dependencies among tasks.

Finally, {\it label()} can be used to group tasks, and to assign the group a common identifier (name), which is in turn used as a reference to implement synchronization at the granularity of task groups (see next).

For example, in lines~\ref{src:mainforbegin}-~\ref{src:mainforend} of Listing~\ref{src:sobel} a separate task is created to compute each row of the output image. The significance of the tasks ranges between 0.1 and 0.9 in a round-robin way (line~\ref{src:tasksig}), which ensures that there will not be extreme, apprehensible quality fluctuations across different areas of the output image. Care has also been taken in this case to avoid using the special values 0.0 and 1.0. Moreover, an approximate version of the task body is implemented by the {\it sbl\_task\_appr} function (lines~\ref{src:sbltaskapprbegin}--\ref{src:sbltaskapprend}). This function implements a light-weight version of the computation, substituting complex arithmetic operations with simpler ones (line~\ref{src:sbltaskapprabs}), while at the same time skipping some filter taps (lines~\ref{src:sbltaskapprskip1}, ~\ref{src:sbltaskapprskip2}). All tasks created in the specific loop belong to the {\it sobel} task group, using {\it img} as input and {\it res} as output (line~\ref{src:taskinoutlabel}).

\begin{center}
\begin{minipage}{\columnwidth}   
\begin{lstlisting}[caption={\#pragma omp taskwait}, label={src:omptaskwait}]
#pragma omp taskwait [on(...)] [label(...)] 
  [ratio(...)]
\end{lstlisting}
\end{minipage}
\end{center}

The proposed programming model supports explicit barrier-type synchronization through the {\it \#pragma omp taskwait} directive (Listing~\ref{src:omptaskwait}). A {\it taskwait} can serve as a global barrier, instructing the runtime to wait for all tasks spawned up to that point in the code. Alternatively, it can implement a barrier at the granularity of a specific task group, if the {\it label()} clause is present; in this case the runtime system waits for the termination of all tasks of that group. Finally, the {\it on()} clause can be used to instruct the runtime to wait for all tasks that affect a specific variable or data construct.

Furthermore, the {\it omp taskwait} barrier can be used to control the minimum quality of application results. Through the {\it ratio()} clause, the programmer can instruct the runtime to execute (at least) the specified percentage of all tasks -- either globally or in a specific group, depending on the existence of the {\it label()} clause -- in their accurate version, while {\it respecting} task significance (i.e., a more significant task should not be executed approximately, while a less significant task is executed accurately). The ratio takes values in the range [0.0, 1.0] and serves as a single, straightforward knob to enforce a minimum quality in the performance / quality / energy optimization space. Smaller ratios give the runtime more energy reduction opportunities, however at a potential quality penalty.

For example, line~\ref{src:sobeltaskwait} of Listing~\ref{src:sobel} specifies a barrier for the tasks of the {\it sobel} task group. The runtime needs to ensure that at least the 35\% most significant tasks of the group will be executed accurately.

The compiler for the programming model is implemented based on a source-to-source compiler infrastructure~\cite{SCOOP}. It recognizes the pragmas introduced by the programmer and lowers them to corresponding calls of the runtime system discussed in Section~\ref{sec:runtime}.

%% file: runtime.tex
\section{Runtime}\label{sec:runtime}

We demonstrate how to extend existing runtime systems
to support our programming model for approximate computing.
To this end, we extend a task-based parallel runtime system that
implements OpenMP 4.0-style task dependencies~\cite{tzenakis2012bddt}.

Our runtime system is organized as a master/slave work-sharing scheduler. The master
thread starts executing the main program sequentially.
For every task call encountered,
the task is enqueued in a per-worker task queue. Tasks are distributed
across workers in round-robin fashion. Workers select the oldest tasks
from their queues for execution. When a worker's queue runs empty,
the worker may steal tasks from other worker's queues.

The runtime system furthermore implements an efficient mechanism for identifying and
enforcing dependencies between tasks that arise from annotations of
the side effects of tasks with \textit{in(...)} and \textit{out(...)} clauses.
Dependence tracking is however not affected by our approximate computing
programming model. As such, we provide no further details on this feature.

\subsection{Runtime API Extension}
The runtime exposes an API that matches with the
pragma-based programming model. Every pragma in the program
is translated in one or more runtime calls.
The runtime API is extended to convey the new information in the programming
model. Task creation is extended to indicate the \emph{task group}
and \emph{significance} of the task, as well as an alternative (approximate)
task function.
On the first use of a task group, the compiler inserts a call
to \textit{tpc\_init\_group()} to create support data structures in the
runtime for the task group. This API call also conveys the per-group
ratio of tasks that must be executed accurately.

An additional waiting API call is created.
Next to the API call \textit{tpc\_wait\_all()},
which waits for all tasks to finish,
we create the API call \textit{tpc\_wait\_group()} to synchronize on the
completion of a task group.


\subsection{Runtime Support for Approximate Computing}
The job of the runtime system is to selectively execute a subset
of the tasks approximately while respecting the constraints given
by the programmer.
The relevant information consists of (i)~the significance of
each task, (ii)~the group a task belongs to, and (iii)~the
fraction of tasks that may be executed approximately for each task group.
Moreover, preference should be given to approximating tasks with
lower significance values as opposed to tasks with high
significance values.

The runtime system has no a priori information on how many tasks will be issued
in a task group, nor what the distribution is of the significance levels
in each task group. This information must be collected at runtime.
In the ideal case, the runtime system knows this information in advance.
Then, it is straightforward to execute approximately those tasks with
the lowest significance in each task group. The policies we design
must however work without this information, and estimate it at runtime.
We define two policies, one globally controlled policy based on buffering
issued tasks and analyzing their properties,
and a policy that estimates the distribution
of significance levels using per-worker local information.

\subsection{Global Task Buffering (GTB)}
\label{sec:overhead}
\begin{figure}[t!]
\begin{lstlisting}[caption={Global task buffering policy to choose the accuracy of a task}, label={src:window}, language=C, basicstyle=\ttfamily\scriptsize]
TaskDesc buffer[BUFFER_SIZE]; // to analyze tasks
size_t task_count = 0;        // buffer occupation
float group_ratio;            // set by #pragma

void buffer_task(TaskDesc t) { // called by master thread
	buffer[task_count] = t;
	task_count++;
	if (task_count == BUFFER_SIZE)
		flush_buffer();
}

void flush_buffer() { // when tasks need to execute
	sort(buffer); // sort by increasing significance
	for (i=0; i<task_count; i++) {
		if (i < group_ratio * task_count)
			issue_accurate_task(buffer[i]);
		else
			issue_approximate_task(buffer[i]);
	}
	task_count = 0;
}
\end{lstlisting}
\end{figure}
In the first policy the master thread buffers a number of tasks
as it creates them, postponing the issue of the tasks in the worker queues.
When the buffer is full, or when the a call to \textit{tpc\_wait\_all()}
or \textit{tpc\_wait\_group()}
is made, the tasks in the buffer are analyzed and sorted by significance.
Given a per-group ratio of accurate tasks $R_g$, and a number of $B$ tasks
in the buffer, then the $R_g\cdot B$ tasks with the highest
significance level are executed accurately.
The tasks are subsequently issued to the worker queues.
The policy is described in Listing~\ref{src:window} for a single task group.
The variables described (buffer, task count and per-group accuracy ratio)
are replicated over all task groups introduced by the programmer.

The task buffering policy is parameterized by the task buffer size.
A larger buffer size allows the runtime to take more informed
decisions. Notably, if the buffer size is sufficiently large,
the runtime can end up buffering all tasks until the corresponding
synchronization barrier is encountered, and thus take a fully
correct decision as to which tasks to run accurately/approximately. 
In our implementation, the buffer size is a configurable parameter
passed to the runtime system at compile time.

The global buffer policy has the potential disadvantage that
it slows done the program by postponing task execution until the buffer is full and sorted.
In the extreme case, the runtime system needs to wait for all tasks  to be issued
and sorted in the buffer before starting their execution. This overhead can be mitigated by using a 
smaller window size and tasks of coarse enough granularity, so that the runtime system 
can overlap task issue with task execution. Using smaller window sizes
will incur the cost of not making fully correct decisions for approximate execution.
Section~\ref{sec:eval} demonstrates that the GTB policy sustains low overhead
in practice.

\subsection{Local Queue History (LQH)}
\label{sec:overhead2}
The local queue history policy avoids the 
step of task buffering.
Tasks are issued to worker queues immediately as they are created. 
The worker decides whether to approximate a task right before
it starts its execution, based on the distribution of significance
levels of the tasks executed so far, and the target ratio
of accurate tasks (supplied by the programmer). 
Hereto, the workers track the number of tasks at each
significance level as they are executed.

Formally, the local queue history policy operates as follows.
Let $t_g(s)$ indicate the number of tasks in task group $g$
observed by a worker with significance $s$ or less.
These statistics are updated for every executed task.
Note that the significance levels $s$ are constrained to the range 0.0 to 1.0.
In the runtime system, we implement 101 discrete (integer) levels to simplify the implementation,
ranging from 0.0 to 1.0 (inclusive) in steps of 0.01.
By construction, $t_g(1.0)$ equals the total number of tasks executed so far.
Let $R_g$ be the target ratio of tasks that should be executed
accurately in task group $g$, as set by the programmer.
Then, assuming a task has significance level $s$, it is
executed accurately if $t_g(s)>(1-R_g)t_g(1.0)$, otherwise it is executed
approximately. 

This policy attempts to achieve a ratio of accurately executed tasks
that converges to $R_g$ and also approximates those tasks with the
lowest significance level, as stipulated by the programming model.

The local queue history algorithm is performed independently by each worker
using only local information from the tasks that appear in their work queue.
Tasks of one group are distributed among the workers via pushing of tasks to different
local queues by the master and work-stealing.  As a result, each worker has only partial information 
about each group.

The overhead of the local queue history algorithm is the bookkeeping of the 
statistics that form the execution history of a group. This happens every 
time a task is executed. Updating statistics includes accessing an array
of size equal to the number of distinct significance levels (101 in the runtime), which
is negligible compared to the granularity of the task.

The local queue history algorithm requires no global snapshot of all tasks in the
program and no synchronization between workers and the master. It is thus more realistic and scalable than 
global task buffering.  However, given that each worker has only a localized view of the tasks issued,
the runtime system can only approximately enforce the quality requirements set by the programmer. 

%% file: experiments.tex
\section{Experimental Evaluation}\label{sec:eval}
We performed a set of experiments to investigate the performance of the proposed programming model and runtime policies, using different benchmark codes that were re-written using the task-based pragma directives. In particular, we evaluate our approach in terms of: 
\begin{inparaenum}[(i)]
	\item The potential for performance and energy reduction;
    \item The potential to allow graceful quality degradation;
    \item The overhead incurred by the runtime mechanisms.
\end{inparaenum}
In the sequel, we introduce the benchmarks and the overall evaluation approach, and discuss the results achieved for various degrees of approximation under different runtime policies. 

\subsection{Approach}\label{sec:eval_approach}

\begin{table}[tb]
\resizebox{\columnwidth}{!}{%
\begin{tabular}{|c|c|c|c|c|c|}
\hline
\multirow{2}{*}{Benchmark} & {\bf A}pproximate & \multicolumn{3}{|c|}{Approx Degree} &\multirow{2}{*} {Quality} \\ \cline{3-5}
		 & or {\bf D}rop  & Mild & Med & Aggr & \\ 						\hline
{\hfil Sobel} & A 		& 80\% & 30\% & 0\% & PSNR\\ \hline
{\hfil DCT} & D 		& 80\% & 40\% & 10\% & PSNR\\ \hline
{\hfil MC} & D, A 		& 100\% & 80\% & 50\% & Rel. Err.\\ \hline
{\hfil Kmeans} & A 		& 80\% & 60\% & 40\% & Rel. Err.\\ \hline
{\hfil Jacobi} & D, A 	& $10^{-4}$ & $10^{-3}$ & $10^{-2}$ & Rel. Err.\\ \hline
{\hfil Fluidanimate}& A	& 50\% & 25\% & 12.5\% & Rel. Err.\\ \hline
\end{tabular}
}
\caption{Benchmarks used for the evaluation. For all cases, except Jacobi, the approximation degree is given by the percentage of accurately executed tasks. In Jacobi, it is given by the error tolerance in convergence of the accurately executed iterations/tasks ($10^{-5}$ in the native version).} 
\label{tab:benchmarks}
\end{table}


\begin{figure}[tb]
\centering
\includegraphics[scale=0.4]{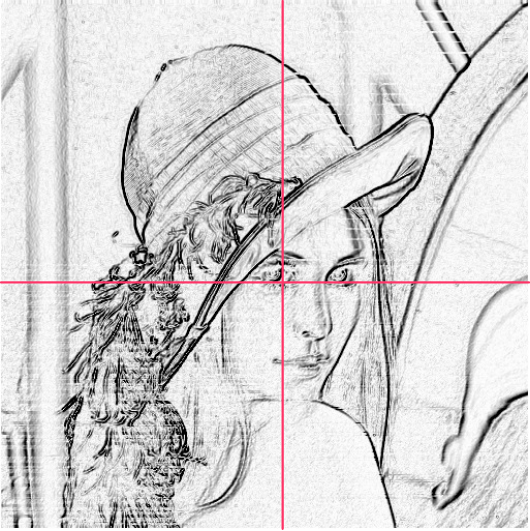}
\caption{ Different levels of approximation for the Sobel benchmark}\label{fig:lena_sobel}
\end{figure}

We use a set of six benchmarks, outlined in Table~\ref{tab:benchmarks}, where we apply different approximation approaches, subject to the nature/characteristics of the respective computation. 

\textit{Sobel} is a 2D filter used for edge detection in images. The approximate version of the tasks uses a lightweight Sobel stencil with just 2/3 of the filter taps. Additionally, it substitutes the costly formula $\sqrt{{sbl_x}^2+{sbl_y^2}}$ with its approximate counterpart $|sbl_x|+|sbl_y|$. The way of assigning significance to tasks ensures that the approximated pixels are uniformly spread throughout the output image.

Discrete Cosine Transform (\textit{DCT}) is a module of the JPEG compression and decompression ~\cite{skodras2001jpeg} algorithm. We assign higher significance to tasks that compute lower frequency coefficients. 

\textit{MC}~\cite{Vavalis_hpde} applies a Monte Carlo approach to estimate the boundary of a subdomain within a larger partial differential equation (PDE) domain, by performing random walks from points of the subdomain boundary to the boundary of the initial domain. Approximate configurations drop a percentage of the random walks and the corresponding computations. A modified, more lightweight methodology is used to decide how far from the current location the next step of a random walk should be.

\textit{K-means} clustering aims to partition \textit{n} observations in a multi-dimensional space into \textit{k} clusters by minimizing the distance of cluster members to a cluster representative. In each iteration the algorithm spawns a number of tasks, each being responsible for a subset of the entire problem. All tasks are assigned the same significance value. The degree of approximation is controlled by the {\it ratio} used at {\it taskwait} pragmas. Approximated tasks compute a simpler version of the euclidean distance, while at the same time considering only a subset (1/8) of the dimensions. Only accurate results are considered when evaluating the convergence criteria. 

\textit{Jacobi} is an iterative solver of diagonally dominant systems of linear equations. We execute the first 5 iterations approximately, by dropping the tasks (and computations) corresponding to the upper right and lower left areas of the matrix. This is not catastrophic, due to the fact that the matrix is diagonally dominant and thus most of the information is within a band near the diagonal. All the following steps, until convergence, are executed accurately, however at a higher target error tolerance than the native execution (see Table~\ref{tab:benchmarks}).

\textit{Fluidanimate}, a code from the PARSEC benchmark suite~\cite{Bienia:2008:PBS:1454115.1454128}, applies the smoothed particle hydrodynamics (SPH) method to compute the movement of a fluid in consecutive time steps. The fluid is represented as a number of particles embedded in a grid. Each time step is executed as either fully accurate or fully approximate, by setting the {\it ratio} clause of the {\it omp taskwait} pragma to either 0.0 or 1.0. In the approximate execution, the new position of each particle is estimated assuming it will move linearly, in the same direction and with the same velocity as it did in the previous time steps. 


Three different degrees of approximation are studied for each benchmark: \textit{Mild, Medium, } and \textit{Aggressive} (see Table~\ref{tab:benchmarks}). They correspond to different choices in the quality vs. energy and performance space. No approximate execution led to abnormal program termination. It should be noted that, with the partial exception of Jacobi, quality control is possible solely by changing the {\it ratio} parameter of the {\it taskwait} pragma. This is indicative of the flexibility of our programming model. As an example, Figure~\ref{fig:lena_sobel} visualizes the results of different degrees of approximation for \textit{Sobel}: the upper left quadrant is computed with no approximation, the upper right is computed with \textit{Mild} approximation, the lower left with \textit{Medium} approximation, whereas the lower right corner is produced when using \textit{Aggressive} approximation.

The quality of the final result is evaluated by comparing it to the output produced by a fully accurate execution of the respective code. The appropriate metric for the quality of the final result differs according to the computation. For benchmarks involving image processing (\textit{DCT, Sobel}), we use the peak signal to noise ratio (\textit{PSNR}) metric, whereas for \textit{MC, Kmeans, Jacobi} and \textit{Fluidanimate} we use the relative error.

In the experiments, we measure the performance of our approach for the different benchmarks and approximation degrees, for the two different runtime policies GTB and LQH. For GTB, we investigate two cases: the buffer size is set so that tasks are buffered until the synchronization barrier (referred to as Max Buffer GTB); the buffer size is set to a smaller value, depending on the computation, so that task execution can start earlier (referred to as GTB). 

As a reference, we compare
our approach against:
\begin{itemize}
	\item A fully accurate execution of each application, using a significance agnostic version of the runtime system. 
    \item An execution using loop perforation~\cite{Sidiroglou-Douskos:2011:MPV:2025113.2025133}, a simple yet usually effective compiler technique for approximation. Loop perforation is also applied in three different degrees of aggressiveness. The perforated version executes the same number of tasks as those executed accurately by our approach.
\end{itemize}

The experimental evaluation is carried out on a system equipped with 2 \textit{Intel(R) Xeon(R) CPU E5-2650} processors clocked at 2.00 GHz, with 64 GB shared memory. Each CPU consists of 8 cores. Although cores support SMT execution (hyper-threading), we deactivated this feature during our experiments. We use Centos 6.5 Linux Operating system with the  2.6.32 Linux kernel. Each execution pinned 16 threads on all 16 cores.

Finally the energy and power are measured using likwid~\cite{treibig2010likwid} to access the Running Average Power Limit (RAPL) registers of the processors.

\subsection{Experimental Results}

\begin{figure*}
\begin{tabular}{ccccc}
& Execution time (secs) & Energy (Joules) & Quality \\
&  lower is better & lower is better & lower is better \\
\rotatebox{90}{\hspace{0.5in}Sobel} & 
\includegraphics[width=0.3\textwidth]{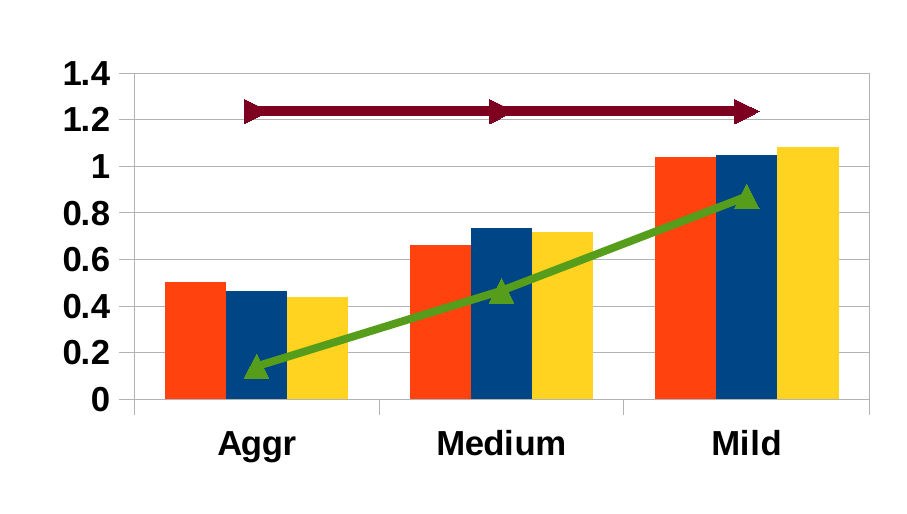}  &
\includegraphics[width=0.3\textwidth]{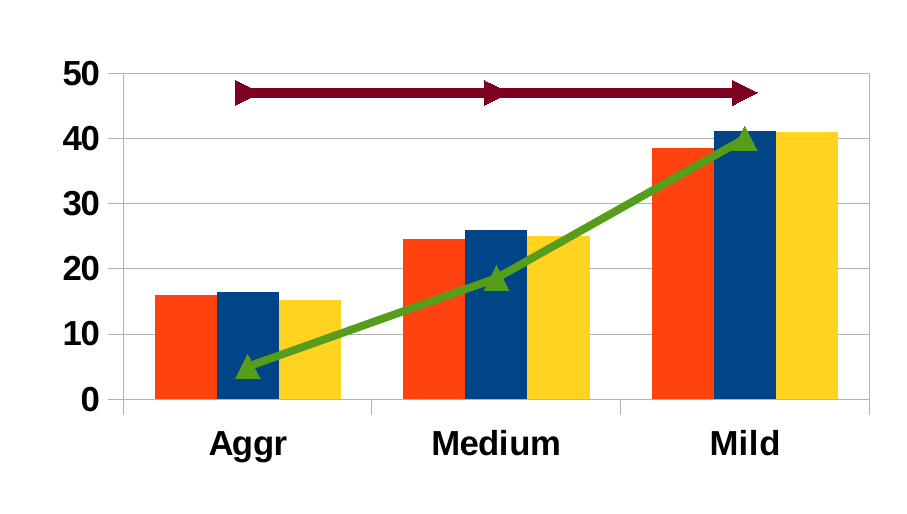}  &
\includegraphics[width=0.3\textwidth]{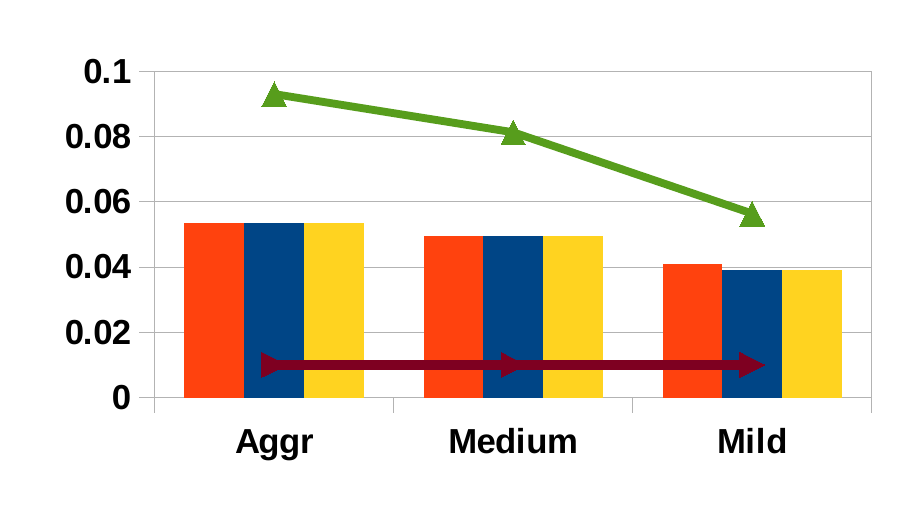} &
\hspace{-0.22in}\rotatebox{90}{\hspace{0.4in}$PSNR^{-1}$} \\

\rotatebox{90}{\hspace{0.5in}DCT} & 
\includegraphics[width=0.3\textwidth]{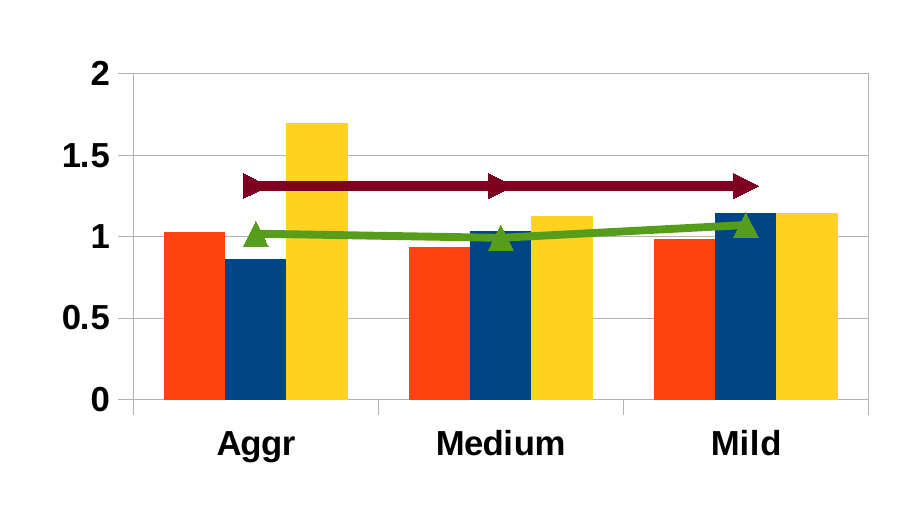}  &
\includegraphics[width=0.3\textwidth]{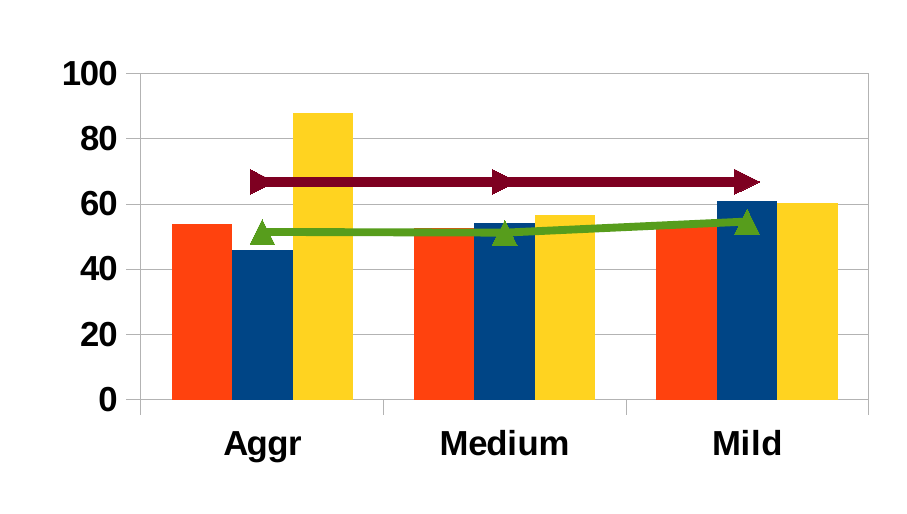}  &
\includegraphics[width=0.3\textwidth]{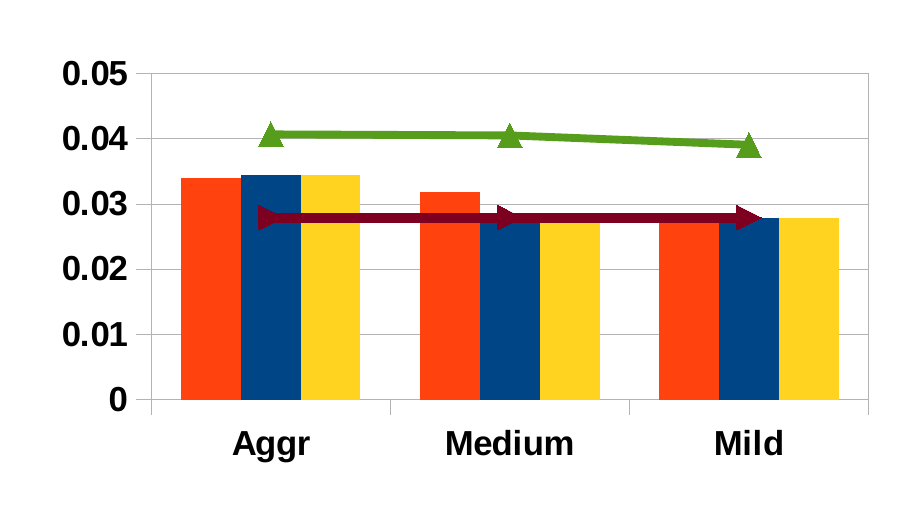} &
\hspace{-0.22in}\rotatebox{90}{\hspace{0.4in}$PSNR^{-1}$} \\

\rotatebox{90}{\hspace{0.5in}MC} & 
\includegraphics[width=0.3\textwidth]{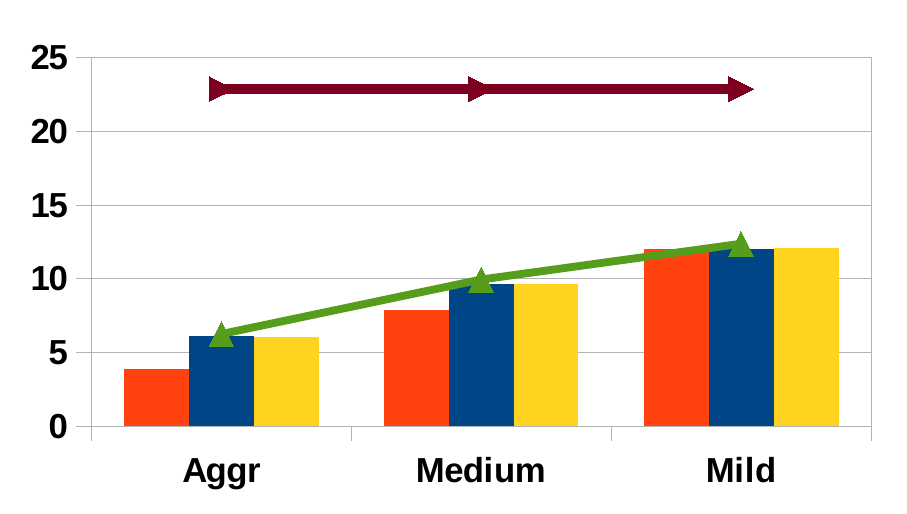}  &
\includegraphics[width=0.3\textwidth]{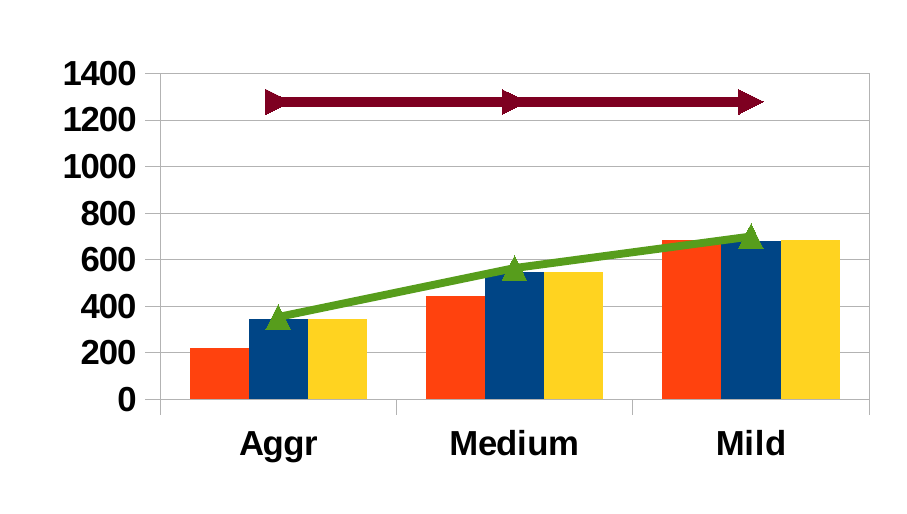}  &
\includegraphics[width=0.3\textwidth]{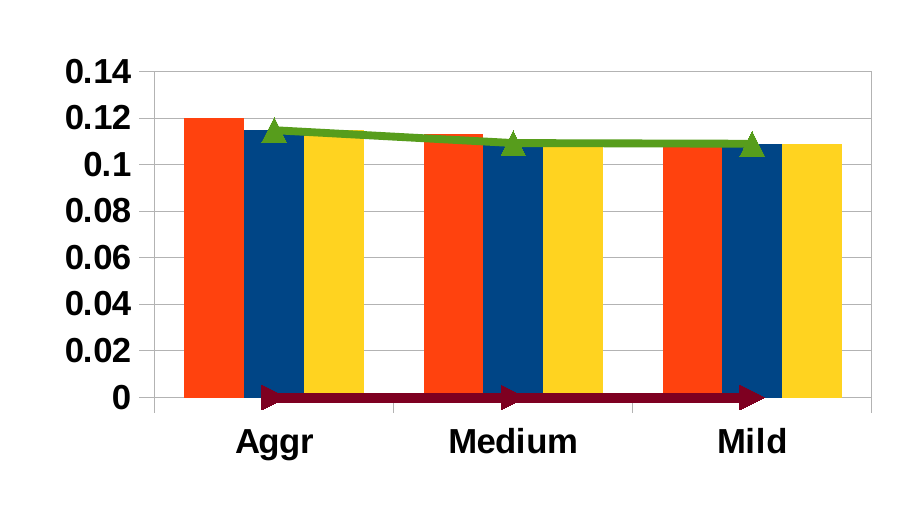} &
\hspace{-0.22in}\rotatebox{90}{\hspace{0.4in}Rel.Error}  \\

\rotatebox{90}{\hspace{0.5in}Kmeans} & 
\includegraphics[width=0.3\textwidth]{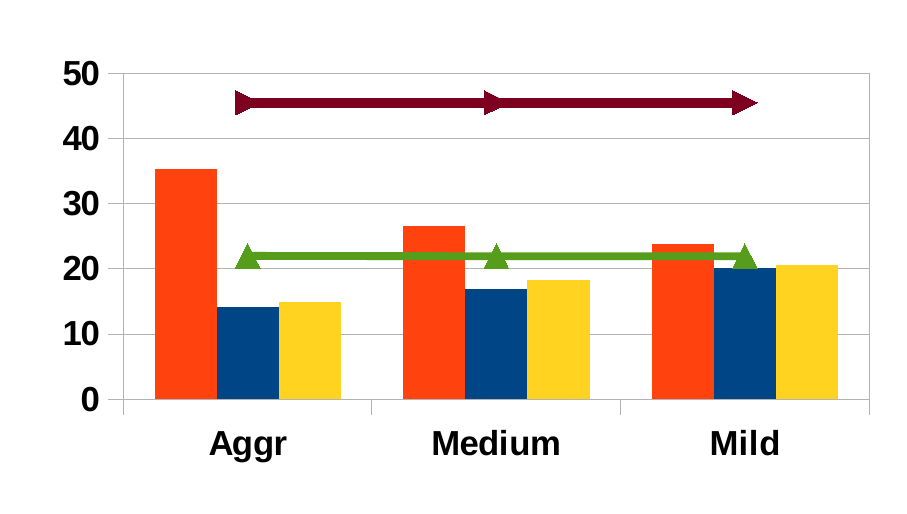}  &
\includegraphics[width=0.3\textwidth]{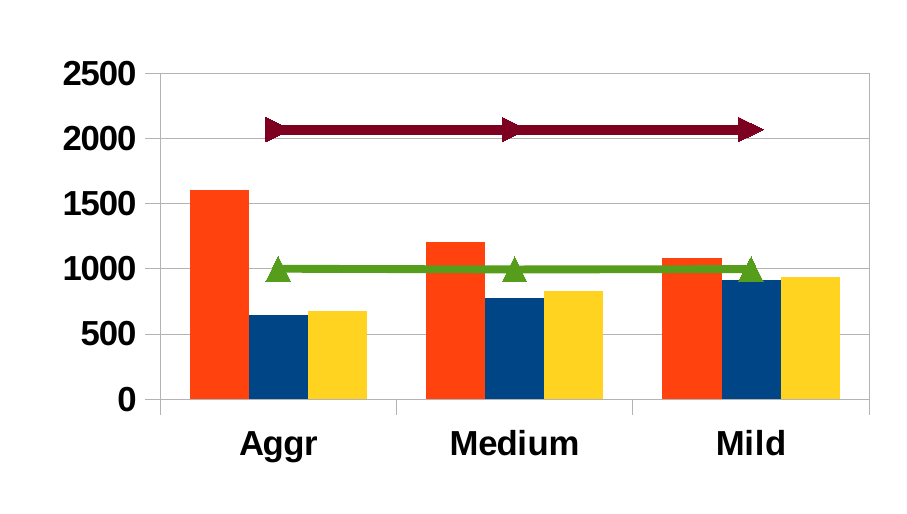}  &
\includegraphics[width=0.3\textwidth]{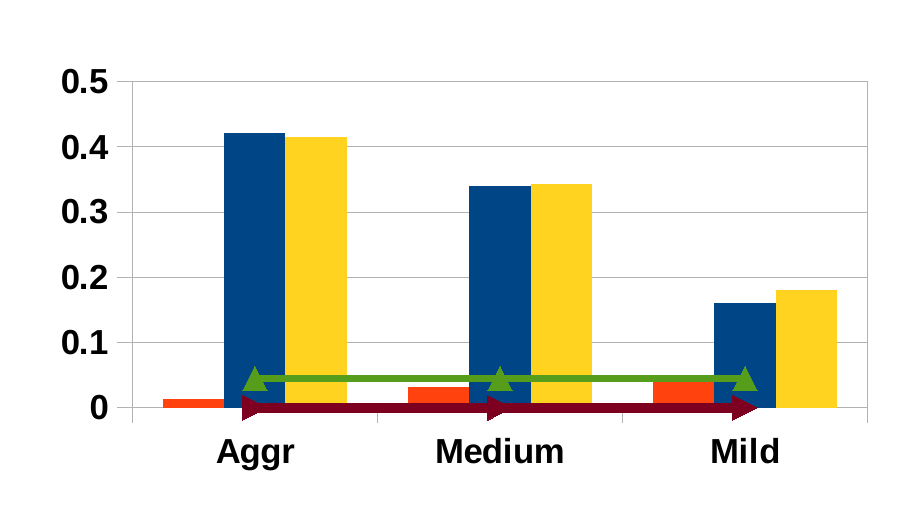} &
\hspace{-0.22in}\rotatebox{90}{\hspace{0.4in}Rel.Error}  \\

\rotatebox{90}{\hspace{0.5in}Jacobi} & 
\includegraphics[width=0.3\textwidth]{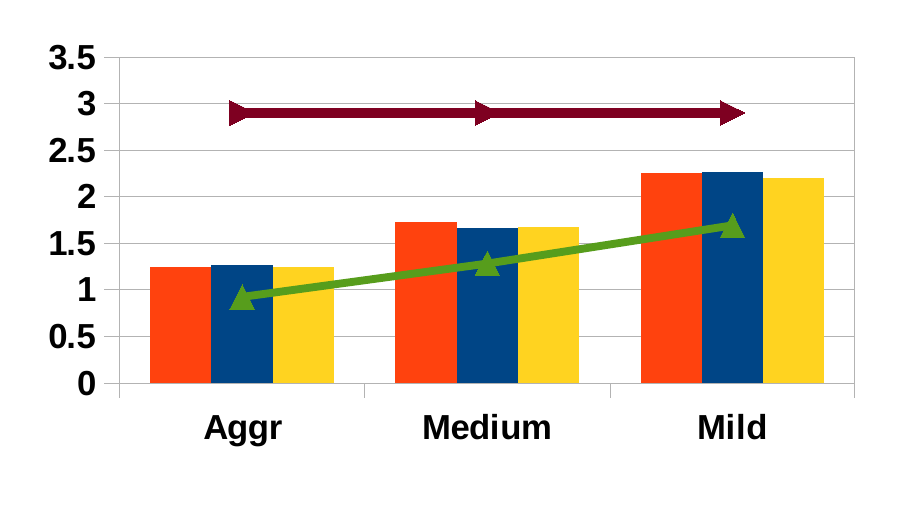}  &
\includegraphics[width=0.3\textwidth]{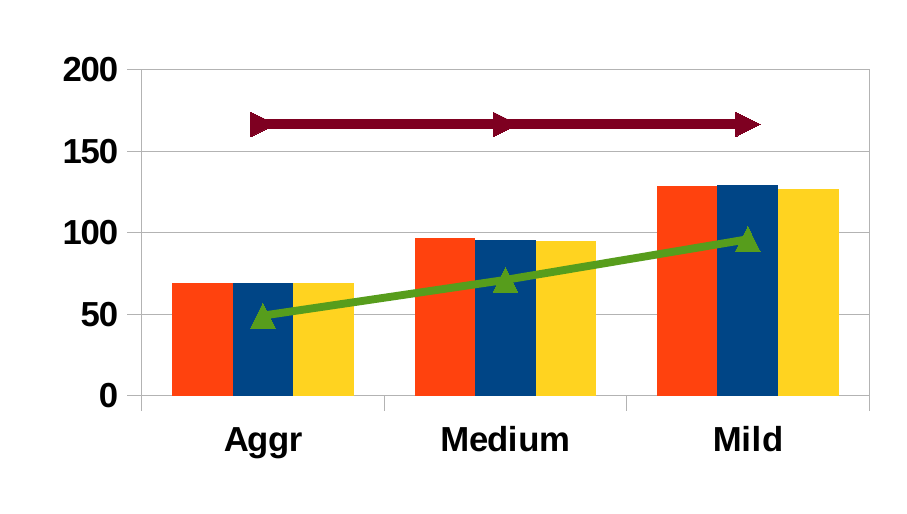}  &
\includegraphics[width=0.3\textwidth]{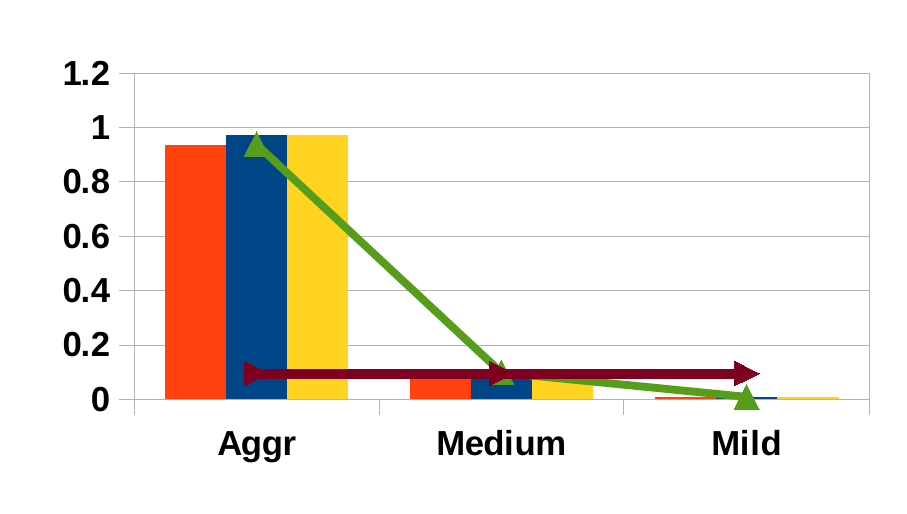} &
\hspace{-0.22in}\rotatebox{90}{\hspace{0.4in}Rel.Error}  \\

\rotatebox{90}{\hspace{0.3in}Fluidanimate} & 
\includegraphics[width=0.3\textwidth]{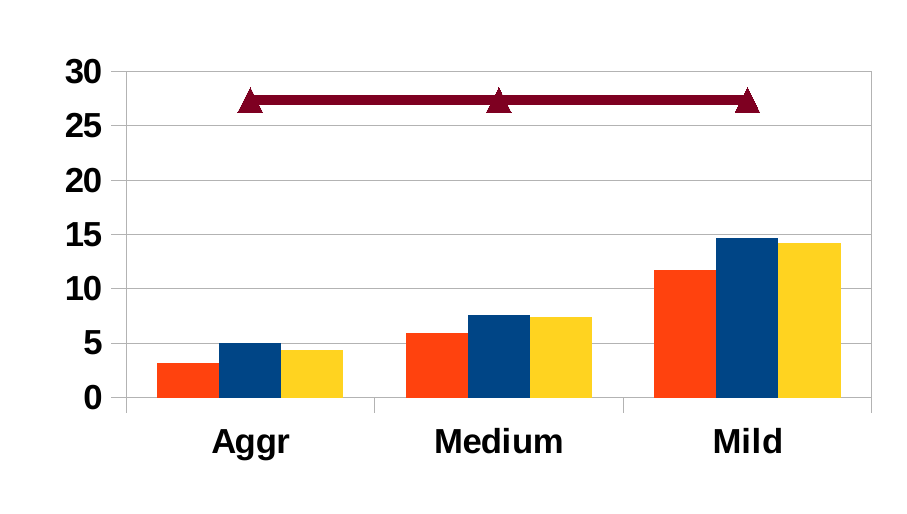}  &
\includegraphics[width=0.3\textwidth]{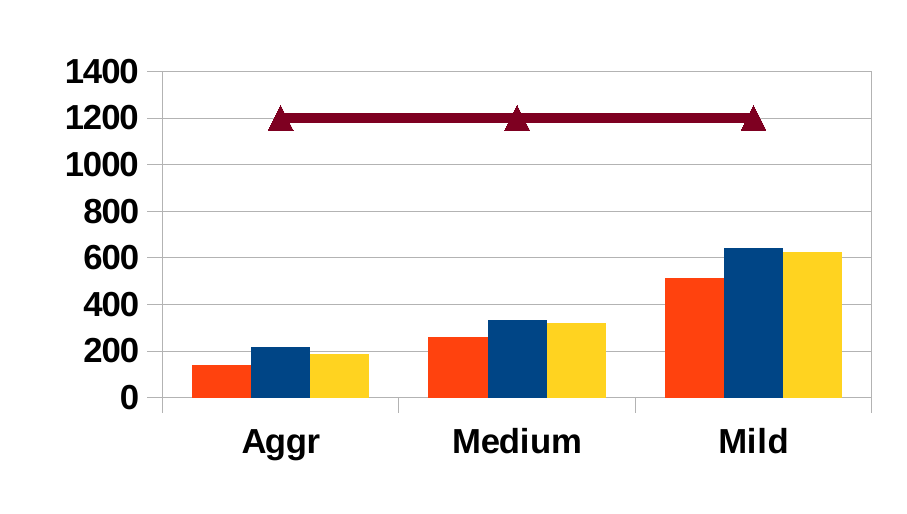}  &
\includegraphics[width=0.3\textwidth]{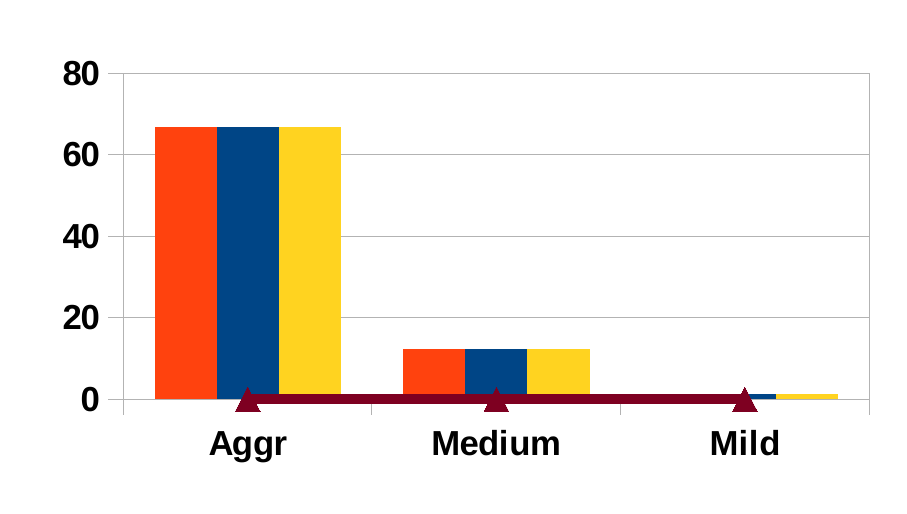}  &
\hspace{-0.22in}\rotatebox{90}{\hspace{0.4in}Rel.Error} \\

 & \multicolumn{3}{c}{\includegraphics[width=0.7\textwidth]{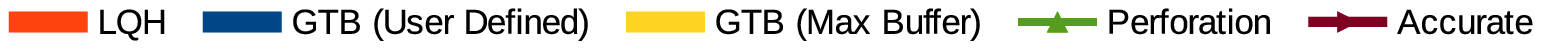} }\\

\end{tabular}
\caption{Execution time, energy and quality of results for the benchmarks used in the experimental evaluation under different runtime policies and degrees of approximation. In all cases lower is better. Quality is depicted as PSNR$^{-1}$ for Sobel and DCT, relative error (\%) is used in all others benchmarks. The accurate execution and the approximate execution using perforation are visualized as lines. Note that perforation was not applicable for Fluidanimate.}
\label{fig:exp_results}
\end{figure*}

Figure \ref{fig:exp_results} depicts the results of the experimental evaluation of our system.  For each benchmark we present execution time, energy consumption and the corresponding error metric. 

\begin{figure}[tb]
\centering
\includegraphics[scale=0.4]{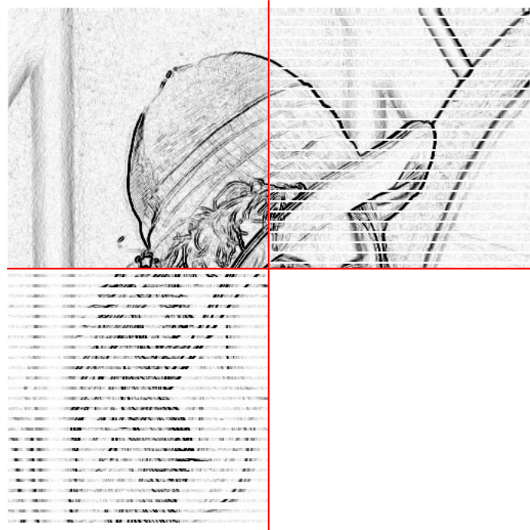}
\caption{ Different levels of perforation for the Sobel benchmark. Accurate execution, Perforation of 20\%, 70\% and 100\% of loop iterations on the upper left, upper right, lower left and lower right quadrants respectively.}\label{fig:lena_perf}
\end{figure}

The approximated versions of the benchmarks execute significantly faster and with less energy 
consumption compared to their accurate counterparts. Although the quality of the application output deteriorates as the approximation level increases, this is typically done in a graceful manner, as it can be observed in Figure~\ref{fig:lena_sobel} and the '{\it Quality}' column of Figure~\ref{fig:exp_results}.

The GTB policies with different buffer sizes are comparable with each other. Even though Max buffer GTB postpones task issue until the creation of all tasks in the group, this does not seem to penalize the policy. In most applications tasks are coarse-grained and are organized in relatively small groups, thus minimizing the task creation overhead and the latency for the creation of all tasks within a group. 
LQH is typically faster and more energy-efficient than both GTB flavors, except for \textit{Kmeans}. 

In the case of \textit{Sobel}, the perforated version seems to significantly outperform our approach in terms of both energy consumption and execution time. However the cost of doing so is unacceptable output quality, even for the mild approximation level as shown in Figure~\ref{fig:lena_perf}. Our programming model and runtime policies achieve graceful quality degradation, resulting in acceptable output even with
aggressive approximation, as illustrated in Figure~\ref{fig:lena_sobel}.

\textit{DCT} is friendly to approximations: it produces visually acceptable results even if a large percentage of the computations is dropped. Our policies, with the exception of the Max Buffer version of GTB, perform comparably to loop perforation in terms of performance and energy consumption, yet resulting in higher quality results\footnote{Note that PSNR is a logarithmic metric}. This is due to the fact that our model offers more flexibility than perforation in defining the relative significance of code regions in DCT. The problematic performance of GTB(Max Buffer) is discussed later in this Section, when evaluating the overhead of the runtime policies and mechanisms.

The approximate version of \textit{MC} significantly outperforms the original accurate version, without suffering much of a penalty on its output quality. Randomized algorithms are inherently susceptible to approximations without requiring much sophistication. It is characteristic that the performance of our approach is almost identical to that of blind loop perforation. We observe that the LQH policy attains slightly better results. In this case, we found that the LQH policy undershoots the requested ratio, evidently executing fewer tasks \footnote{4.6\% and 5.1\% more that requested tasks are approximated for the aggressive and the medium case respectively.}. This affects quality, which is lower than that achieved by the rest of the policies.

\textit{Kmeans} behaves gracefully as the level of approximation increases. Even in the aggressive case, all policies demonstrate relative errors less than 0.45\%. The GTB policies are superior in terms of 
execution time and energy consumption in comparison with the perforated version of the benchmark. 
Noticeably, the LQH policy exhibits slow convergence to the termination criteria. The application terminates when the number of objects which move to another cluster is less than 1/1000 of the total object population. As mentioned in the Section~\ref{sec:eval_approach}, objects which are computed approximately do not participate in the termination criteria. GTB policies behave deterministically, therefore always selecting tasks corresponding to specific objects for accurate executions. On the other hand, due to the effects dynamic load balancing in the runtime and its localized perspective, LQH tends to evaluate accurately different objects in each iteration. Therefore, it is more challenging for LQH to achieve the termination criterion. Nevertheless, LQH produces results with the same quality as a fully accurate execution with significant performance and energy benefits.

\textit{Jacobi} is a particular application, in the sense that approximations can affect its rate of convergence in deterministic, yet hard to predict and analyze ways. The blind perforation version requires fewer  iterations to converge, thus resulting in lower energy consumption than our policies. Interestingly enough, it also results in a solution closer to the real one, compared with the accurate execution.

The perforation mechanism could not be applied on top of the \textit{Fluidanimate} benchmark. This is because if the evaluation of the movement of part of the particles during a time-step is totally dropped, the physics of the fluid are violated leading to completely wrong results. Our programming model offers the programmer the expressiveness to approximate the movement of the liquid for a set of time-steps. Moreover, in order to ensure stability, in is necessary to alternate accurate and approximate time steps. In our programming model this is achieved in a trivial manner, by alternating the parameter of the {\it ratio} clause at {\it taskbarrier} pragmas between 100\% and the desired value in consecutive time steps. It is worth noting that \textit{Fluidanimate} is so sensitive to errors that only the mild degree of approximation leads to acceptable results. Even so, the LQH policy requires less than half the energy of the accurate execution, with the 2 versions of the GTB policy being almost as efficient. 

\begin{figure}[tb]
\centering
\includegraphics[width=0.45\textwidth]{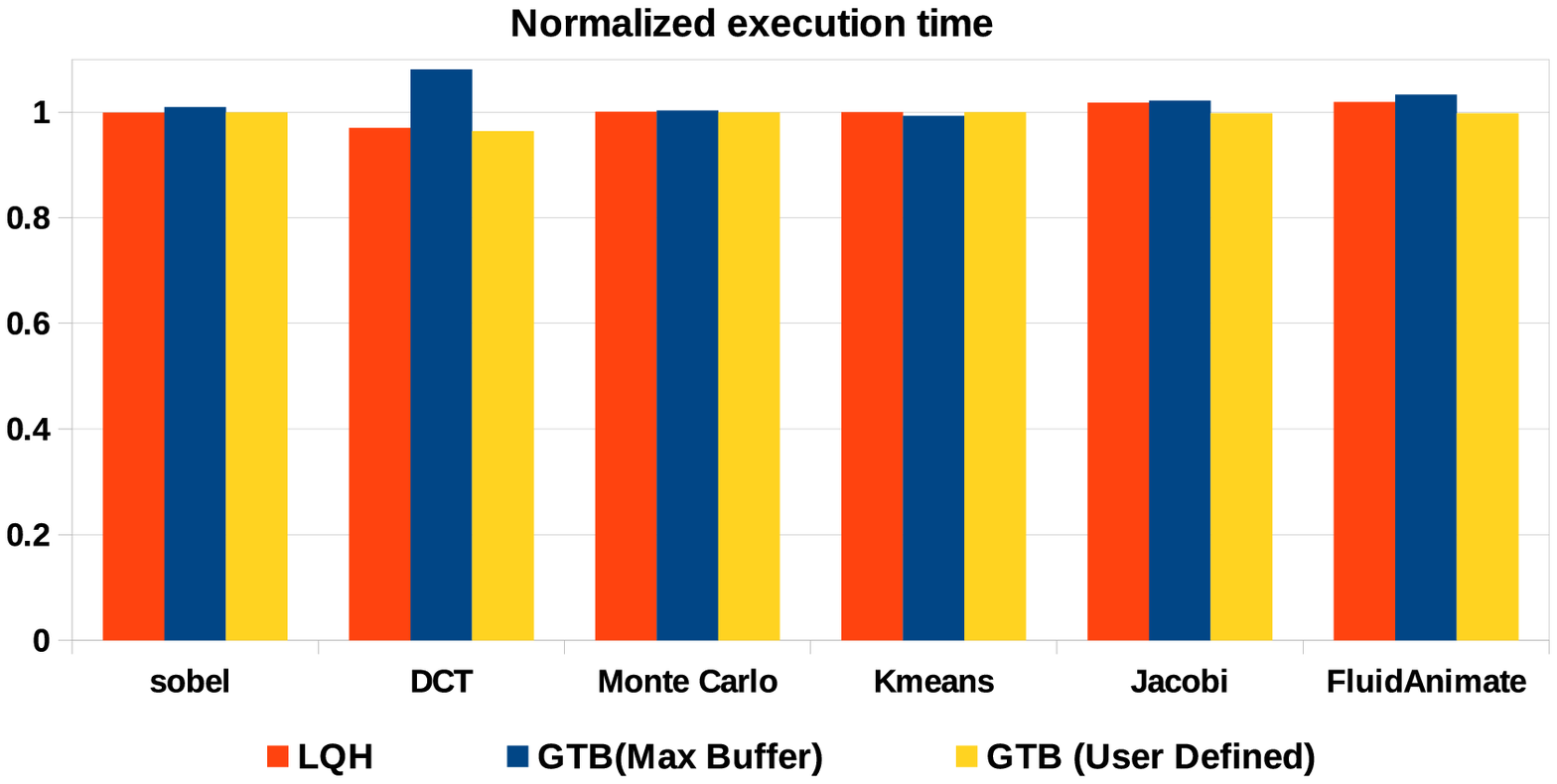}
\caption{The normalized execution time of benchmarks under different task categorization policies, with respect to that over the significance-agnostic runtime system}\label{fig:execution_times}
\end{figure}

Following, we evaluate the overhead of the runtime policies and mechanisms discussed in Sections~\ref{sec:overhead} and \ref{sec:overhead2}. We measure the performance of each benchmark when executed with a significance-agnostic version of the runtime system, which does not include the execution paths for classifying and executing tasks according to significance. We then compare it with the performance attained when executing the benchmarks with the significance-aware version of the runtime. All tasks are created with the same significance and the ratio of tasks executed accurately is set to 100\%, therefore eliminating any benefits of approximate execution. Figure~\ref{fig:execution_times} summarizes the results. It is evident that the significance-aware runtime system typically incurs negligible overhead. The overhead reaches in the order of 7\% in the worst case (DCT under the GTB Max Buffer policy). DCT creates many lightweight tasks, therefore stressing the runtime. At the same time, given that for DCT task creation is a non-negligible percentage of the total execution time, the latency between task creation and task issue introduced by the Max Buffer version of the GTB policy results in a measurable overhead.

\begin{table}[tb]
\resizebox{\columnwidth}{!}{%
\begin{tabular}{|c|c|c|c|c|c|c|}
\hline
\multirow{2}{*}{Benchmark} & \multicolumn{3}{|c|}{(\%) Inversed Significance Tasks}  & \multicolumn{3}{|c|}{Average Ratio Diff}  \\ \cline{2-7}
  & LQH & GTB(UD) & GTB (MB)& LQH & GTB(UD) & GTB (MB) \\ \hline
  Sobel &  2.7  &  0 &  0   & 0.07&  0   &  0\\ \hline
  DCT  &  2.7  &  0 &  0   & 0.18    &  0   & 0 \\ \hline
  MC   &  4.8  &  0 &  0   & 0.17  & 0    &0  \\ \hline
  KMeans  &  0   & 0 &  0   & 0.9  &  0  &  0\\ \hline
  Jacobi  &  0  &  0 &  0   & 0.12 &0     &  0\\ \hline
  FluidAnimate  &  0  &  0 &  0  & 0 &  0   & 0 \\ \hline
\end{tabular}
}
\caption{Degree of accuracy of the proposed policies.}\label{tab:cat_policies}
\end{table}

The last step of our evaluation focuses on the accuracy of the policies in terms of respecting the significance of tasks and the user-supplied ratio of accurate tasks to be executed. Table~\ref{tab:cat_policies} summarizes the results. The average offset in the ratio of accurate tasks executed is calculated by the following formula:\\
\begin{center}
$ratio\_diff= \frac{\sum_{i=1}^{Groups}{| requested ratio_i - provided ratio_i |}}{Total Groups}  $
\end{center}

The two versions of GTB respect perfectly task significance and the user-specified ratio. This is totally expected for the Max Window version of GTB. The version of GTB using a limited window benefits by the relatively small task groups created by the applications and the smoothly distributed significance values in tasks of each group. LQH, in turn, is inherently more inaccurate, due to its localized perspective. It manages to avoid significance inversion only in cases where all tasks within each task group have the same significance ({\it Kmeans, Jacobi, Fluidanimate}). Even in these cases, LQH may slightly deviate from the specified ratio, due to the loose collaboration of policy modules active on different workers. 

%% file: related.tex
\section{Related Work}
\label{sec:related_work}

We classify related work in approximate computation into general-purpose frameworks, parallel programming and execution models that implement approximate 
computation, and other approaches, including domain-specific frameworks and hardware support for approximate computation.  Finally, we review prior work on runtime energy optimization of parallel programs, that does
not employ approximation.

\subsection{General-Purpose Approximation Frameworks}

Several frameworks for approximate computing discard parts of code at runtime, while asserting that the quality of the result complies with 
quality criteria provided by the programmer. 
Green~\cite{Baek:2010:GFS:1806596.1806620} is an API for loop-level and function approximation. 
Loops are approximated with a reduction of the loop trip count. Functions are approximated
with multi-versioning. The API includes 
calibration functions that build application-specific QoS models for the outputs of 
the approximated blocks of code, as well as re-calibration functions for correcting unacceptable errors that may incur due to 
approximation.  Sloan et al.~\cite{Sloan:2012:SDS:2228360.2228524} provide guidelines for manual control of approximate computation and error checking in software.  
These frameworks delegate the control of approximate code execution to the programmer. We explore an alternative approach where
the programmer uses a higher level of abstraction for approximation, namely computational significance, while the system software translates this
abstraction into energy- and performance-efficient approximate execution.

Loop perforation~\cite{Sidiroglou-Douskos:2011:MPV:2025113.2025133} is a compiler technique that
classifies loop iterations into critical and non-critical ones. The latter can be dropped, as long as the results of the loop are acceptable from a quality standpoint.  
Input sampling and code versioning~\cite{Zhu:2012:RAP:2103656.2103710}  also use the compiler to selectively discard inputs to functions
and substitute accurate function implementations with approximate ones. Similarly to loop perforation and code versioning,
our framework benefits from task dropping and the execution of approximate versions of tasks. However, we follow a different 
approach whereby these optimizations are driven from user input on the relative significance of code blocks and
are used selectively in the runtime system to meet user-defined quality criteria. 

EnerJ~\cite{Sampson:2011:EAD:1993498.1993518} implements approximate data types and supports user-defined ``approximable'' methods, without tying these
abstractions to a specific approximate execution model.  To achieve energy savings, the prototype implementation of EnerJ uses a simulated environment
where it stores approximate data types in DRAM with low refresh rate and SRAM with low supply voltage. Approximable methods are executed on aggressively voltage-scaled processors, with
ISA extensions for approximation~\cite{Esmaeilzadeh:2012:ASD:2150976.2151008,Sampson:2013:ASS:2540708.2540712}. Similarly to our framework, EnerJ provides
abstractions that allow the programmer to provide hints on where approximate execution can be safely used in a program. Contrary to our framework, EnerJ does not use a runtime substrate 
for approximation on general-purpose hardware and does not consider code dropping or task-parallel execution.  



\subsection{Parallel Approximation Frameworks}

Quickstep~\cite{Misailovic:2013:PSP:2465787.2465790} is a tool that approximately parallelizes sequential programs. The parallelized programs are subjected to
statistical accuracy tests for correctness. Quickstep tolerates races that occur after removing synchronization operations that would otherwise be necessary to preserve the semantics of 
the sequential program.  Quickstep thus exposes additional parallelization and optimization opportunities via approximating the data and control dependencies in a program. On the other
hand, QuickStep does not enable algorithmic and application-specific approximation, which is the focus of our work. 

Variability-aware OpenMP~\cite{Rahimi:2013:VOE:2555692.2555727} is a set of OpenMP extensions that enable a programmer to specify blocks of code that can be computed approximately,
The programmer may also specify error tolerance in terms of the number of most significant bits in a variable which are guaranteed to be correct.
Variability-aware OpenMP applies approximation only to specific FPU operations, which execute on specialized FPUs with configurable accuracy. 
Our framework applies selective approximation at the granularity of tasks, using the significance abstraction. 
Our programming and execution model thus provides additional flexibility to drop or approximate code, while preserving output quality. 
Furthermore, our framework does not require specialized hardware support. 

%
Variation-tolerant OpenMP~\cite{Rahimi:2013:VOT:2485288.2485422} uses a runtime system that
characterizes OpenMP tasks in terms of their vulnerability to errors. The runtime system assesses
error vulnerability of tasks online, similarly to our LQH policy for significance characterization.
The variation-tolerant OpenMP runtime uses a hardware error counter to apportion errors to tasks and estimate task vulnerability to errors. The scheduler is a variant of an FCFS, centralized scheduler that uses task vulnerability to select the cores on which each task runs, in order to minimize the number of instructions that are likely to incur errors. Variation-tolerant OpenMP does not consider explicitly identified approximate code and its selective execution for quality-aware energy and performance optimization. 


\subsection{Other Approximation Frameworks}

Several software and hardware schemes for approximate computing follow a domain-specific approach.  ApproxIt~\cite{Zhang:2014:AAC:2593069.2593092} is a framework for approximate iterative methods, based on a lightweight quality control mechanism. Unlike our task-based approach, ApproxIt uses coarse-grain approximation at
 a minimum granularity of one solver iteration.  Gschwandtner et al. use a similar iterative approach to execute 
error-tolerant solvers on processors that operate with near-threshold voltage (NTC) and reduce energy consumption by replacing cores operating at nominal voltage 
with NTC cores~\cite{enahpc14}.  Schmoll et al.~\cite{Schmoll:2013:IFR:2536747.2536753} present algorithmic and static analysis techniques to detect variables that must be computed reliably and variables that can be computed approximately  in an H.264 video decoder. Although we follow a domain-agnostic approach in our approximate computing framework, we provide
sufficient abstractions for implementing the aforementioned application-specific approximation methods.

SAGE~\cite{Samadi:2013:SSA:2540708.2540711} is a compiler and runtime environment for automatic generation of approximate kernels in machine learning and image processing applications. Paraprox~\cite{Samadi:2014:PPA:2541940.2541948} implements transparent approximation for data-parallel programs by recognizing common algorithmic kernels and replacing them with approximate equivalents.  ASAC~\cite{Roy:2014:AAS:2597809.2597812} provides
sensitivity analysis for automatically generated code annotations that quantify significance.  
We do not explore automatic generation of approximate code in this work. However, our techniques for 
quality-aware, selective execution of approximate code are directly applicable to scenarios where 
the approximate code is derived from a compiler, instead of source code annotations. 

Hardware support for approximate computation has taken the form of programmable vector processors~\cite{Venkataramani:2013:QPV:2540708.2540710}, neural networks that approximate the results of code regions 
in hardware~\cite{Esmaeilzadeh:2012:NAG:2457472.2457519}, and low-voltage probabilistic storage~\cite{Salajegheh:2013:HST:2465787.2465793}. These frameworks assume non-trivial,
architecture-specific support from the system software stack, whereas we depend only on compiler and runtime support for task-parallel execution, which is already widely available on 
commodity multi-core systems. 
ESRA~\cite{Leem:2010:EER:1870926.1871302} is a multi-core architecture where cores are either fully reliable or have relaxed reliability. Programs running on ESRA
divide their code into critical (typically control code) and non-critical (typically data processing code) parts and assign these to reliable or unreliable cores, respectively. 
Therefore, ESRA uses an explicit and application-specific assignment of code to cores with different levels of reliability.  We follow a different approach whereby the
programmer uses significance to implicitly indicate code that can be approximated and the runtime system implements selective approximation. In our framework, accurate and approximate code may run 
on any core for load balancing purposes.

%% file: conclusions.tex
\section{Conclusions}\label{sec:conclusions}

We introduced a programming model that supports approximate computing at the granularity of tasks. Tasks are widely used to express parallelism in a high-level and platform-neutral way. We believe that tasks can also be used to introduce approximate versions of specific parts of the computation in a structured way that is amenable to flexible runtime scheduling to achieve energy-efficient program execution at a controllable degradation of output quality. 

We also introduced extensions to a task-based runtime system to exploit significance information, along with a set of significance-centric scheduling policies for elastically deciding which tasks to execute accurately and which approximately, while at the same time respecting programmer's specifications.

We have performed a first evaluation of our implementation on an Intel-based multiprocessor consisting of twin multicore sockets. The results across several different benchmark codes are encouraging, and show that the programmer can easily target different energy-quality trade-offs, by adjusting in the majority of cases a single parameter: the percentage of tasks to execute accurately.    

In the future, we wish to explore more optimization scenarios, such as DFVS in conjunction with suitable runtime policies for executing approximate (and more light-weight) task versions on the slower but also less power-hungry CPUs, as well as for using more such cores to make up for this slower execution. We are also interested in extending our programming model to support approximate computing on top of ultra low-power but unreliable hardware.